\documentclass[twocolumn]{aastex63}

\usepackage{amsmath,amssymb,bbold,mathrsfs}
\usepackage{textcase}
\usepackage{subfigure}

\usepackage{xcolor}
\usepackage{soul}
\definecolor{g}{RGB}{0,160,0}
\setstcolor{g}
\definecolor{b}{RGB}{0,0,160}
\setstcolor{b}
\definecolor{r}{RGB}{250,0,0}
\setstcolor{r}

\newcommand{\fref}[1]{Fig.~\ref{#1}}
\newcommand{\TAUE}[1]{${\rm log_{10}}(\tau_{500})=#1$}

\usepackage[nonumberlist,nosuper]{glossaries}
\setacronymstyle{long-short}

\newacronym{E1}{E1}{electric dipole}
\newacronym{M2}{M2}{magnetic quadrupole}
\newacronym{MIT}{MIT}{Magnetic field induced transitions}
\newacronym{LS}{LS}{Russell-Saunders}
\newacronym{ZIMPOL}{ZIMPOL-3}{Zurich Imaging Polarimeter-3}
\newacronym{IRSOL}{IRSOL}{Istituto ricerche solari Aldo e Cele Dacc\`o}
\newacronym{RT}{RT}{radiative transfer}
\newacronym{RMS}{RMS}{root mean square}

\begin{document}

\title{Observation and Modeling of the circular polarization \\ of the \ion{Cr}{1} magnetic field induced transition at 533.03~nm}

\author[0000-0001-5612-4457]{Hao\ Li}
\affil{Instituto de Astrof\'{\i}sica de Canarias, E-38205 La Laguna, Tenerife, Spain}
\affil{Departamento de Astrof\'\i sica, Universidad de La Laguna, E-38206 La Laguna, Tenerife, Spain}
\author[0000-0003-1465-5692]{Tanaus\'u\ del Pino Alem\'an}
\affil{Instituto de Astrof\'{\i}sica de Canarias, E-38205 La Laguna, Tenerife, Spain}
\affil{Departamento de Astrof\'\i sica, Universidad de La Laguna, E-38206 La Laguna, Tenerife, Spain}
\author[0000-0001-5131-4139]{Javier\ Trujillo Bueno}
\affil{Instituto de Astrof\'{\i}sica de Canarias, E-38205 La Laguna, Tenerife, Spain}
\affil{Departamento de Astrof\'\i sica, Universidad de La Laguna, E-38206 La Laguna, Tenerife, Spain}
\affil{Consejo Superior de Investigaciones Cient\'{\i}ficas, Spain}
\author[0000-0002-3594-2247]{Franziska Zeuner}
\affil{Istituto ricerche solari Aldo e Cele Dacc\`o (IRSOL), Faculty of Informatics, Università della Svizzera italiana, CH-6605 Locarno, Switzerland}


\begin{abstract}
We study the circular polarization of the magnetic field induced transition (MIT)
between the $3d^5(^6S)4d\ ^7D_2$ and $3d^5(^6S)4p\ ^7P_4^\circ$ states of \ion{Cr}{1}
at 533.03~nm (wavelength in air).
The fractional circular polarization $V/I$ of this spectral
line resulting from the solution of the radiation transfer problem in a sunspot
model permeated by a homogeneous magnetic field of 3~kG shows amplitudes of about
$2\%$. Spectro-polarimetric observations of two sunspots were obtained with the
Zurich Imaging Polarimeter-3 at the \gls*{IRSOL} observatory in Locarno, Switzerland.
The observed $V/I$ profiles show approximately anti-symmetrical shapes with an
amplitude of around $0.1\%$ and $0.2\%$ for the two sunspots. The center of this
profile coincides with the wavelengths predicted for the above-mentioned MIT. We apply an
inversion code to the spectro-polarimetric data of the \ion{Cr}{1} permitted
lines at 532.91 and 532.98~nm, as well as to the MIT line at 533.03~nm,
to infer a stratification of the emitting atmosphere. We compare the $V/I$
profiles synthesized in the inferred atmosphere models with the observations,
showing that the observed signal likely corresponds to the MIT line.
\end{abstract}

\section{Introduction} \label{intro}

\gls*{MIT} are forbidden transitions, i.e., not fulfilling the relevant selection
rules, whose oscillator strength is enhanced due to the state-mixing induced
by the presence of a magnetic field, as the quantum number that made them
forbidden ceases being a good quantum number.
\citet{Beiersdorfer2003PhRvL} identified a \gls*{MIT} of \ion{Ar}{9} in a
laboratory experiment for magnetic fields in excess of 10~kG. Relatively recently,
theoretical investigations were carried out by \citet{Li2013PhRvA} and
\citet{Grumer2014PhyS} to accurately compute the transition rates of
\gls*{MIT}s.

During the last years there has been a growing interest of the \gls*{MIT}s in
solar physics research, in particular in relation to magnetic field inference
in the solar corona. Since the transition rate of a \gls*{MIT} depends on the
degree of mixing of atomic states, this rate is therefore sensitive to the
magnetic field strength. In particular, the \ion{Fe}{10} \gls*{MIT} at
257.26~\AA\ has been proposed as a probe of the magnetic field in the solar
corona (\citealt{Li2015ApJ,Li2016ApJ,Li2021ApJ}, and \citealt{Judge2016ApJ}).
A laboratory experiment by \citet{Xu2022ApJ} has demonstrated the potential of
this line for magnetic field diagnostics. 

The \ion{Fe}{10} \gls*{MIT} at 257.26~\AA\ has been used to infer the coronal
magnetic field strength in active regions 
(\citealt{Si2020ApJ,Landi2020ApJ,Landi2021ApJ}, and \citealt{Brooks2021ApJ}) 
from observations obtained by the EUV
Imaging Spectrometer onboard the Hinode satellite \citep{Culhane2007SoPh}. 
\citet{Chen2021ApJa,Chen2021ApJb,Chen2023MNRAS,MartinezSykora2022ApJ}, and \citet{Liu2022ApJ}
have also demonstrated, using magneto-hydrodynamic models,
that the \ion{Fe}{10} \gls*{MIT} line can be used to determine the
magnetic field strength not only in the solar corona, but also in the
corona of other stars. For a review of both the theory to compute the
transition rates of \gls*{MIT} and the application of the magnetic
inference method to the solar corona, see \citet{Chen2023RAA}.

The above-mentioned investigations focused on the intensity of the considered
\gls*{MIT}. In order to study the polarization of a \gls*{MIT}, and motivated
by the fact that the Sun is a laboratory for atomic physics research, we have
checked the solar visible spectrum in the search of a \gls*{MIT}, of which
the polarization can be observed from ground based solar telescopes.
The transition rate of a \gls*{MIT}, to the first-order perturbation approximation, 
is proportional to $B^2 \lambda^{-3} \Delta E^{-2}$ \citep{Li2015ApJ}, with $B$ the magnetic
field strength, $\lambda$ the wavelength of the transition, 
and $\Delta E$ the energy separation between the mixed states. Therefore,
the smaller the energy separation between mixing states, the larger 
the transition rate of the \gls*{MIT}.

In this work we report on the \ion{Cr}{1} \gls*{MIT} at 533.03~nm,
pertaining to the multiplet with upper term $3d^5(^6S)4d\ ^7D$ and lower
term $3d^5(^6S)4p\ ^7P^\circ$. The fine structure separation within the 
upper term is on the order of $\rm 10^{-4}~eV$. Due to its relatively
small ionization potential, the \ion{Cr}{1} atoms are dominant mainly
in the relatively cold photosphere, where the magnetic field is typically
stronger compared with the upper layers of the solar atmosphere.
Another advantage of this line is that it is spectrally resolvable, i.e., 
it is not blended with permitted lines of the same multiplet (as it is
the case for the already mentioned \ion{Fe}{10} 257.262~\AA\ or
\ion{Ar}{9} lines), due to its relatively long wavelength.
Furthermore, the Zeeman split of this transition is significant.
For all these reasons, the circular polarization of this \gls*{MIT}
can be observed in sunspots from ground-based telescopes. 

In Section 2 of this paper we briefly introduce the atomic model used in
the \gls*{RT} modeling of the \ion{Cr}{1} \gls*{MIT}. The
line profiles synthesized in model C of \citeauthor{Fontenla1993ApJ}
(\citeyear{Fontenla1993ApJ}; hereafter, FAL-C model) and model E 
of \citeauthor{Maltby1986ApJ} (\citeyear{Maltby1986ApJ}; hereafter, M-E model)
are compared with the solar spectral atlas. In Section 3 we show the
observations in two sunspots acquired with the Zurich Imaging Polarimeter-3
\citep[ZIMPOL-3;][]{Ramelli2010SPIE} at the \gls*{IRSOL} located in Locarno, Switzerland.
We apply the HanleRT Tenerife Inversion Code
(HanleRT-TIC; \citealt{Tanausu2016ApJ,Li2022ApJ}) to
invert the data and infer the stratification of the underlying solar
atmosphere. The fractional circular polarization $V/I$ profiles of the
\ion{Cr}{1} \gls*{MIT} synthesized in the inferred model atmosphere
are shown and compared with the observation. Our conclusions are
summarized in Section 4.

\section{Atomic model and spectral synthesis} \label{model}

\begin{figure}[htp]
\center
\includegraphics[width=0.50\textwidth]{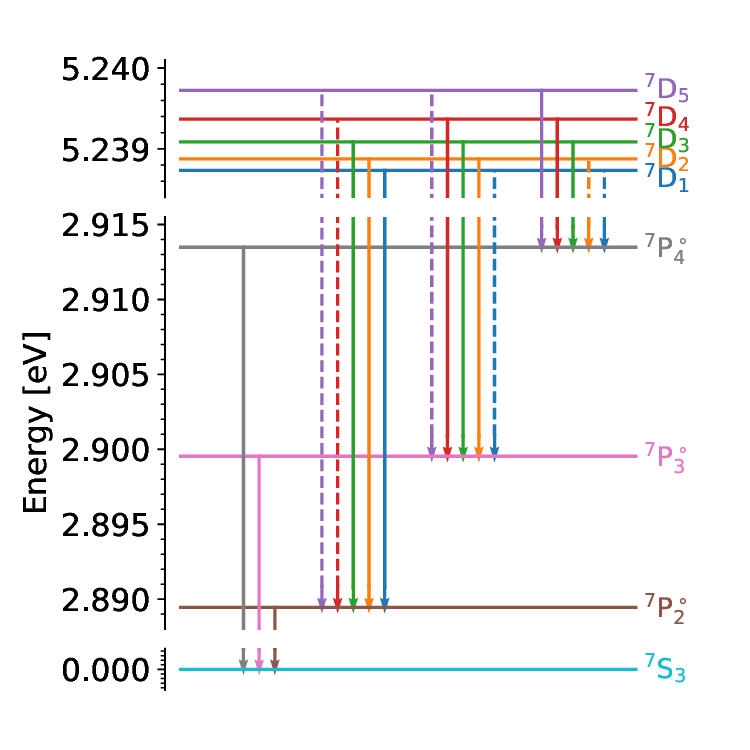}
\caption{Grotrian diagram showing the \ion{Cr}{1} $3d^5(^6S)4s \ ^7S$,
$3d^5(^6S)4p\ ^7P^\circ$, and $3d^5(^6S)4d\ ^7D$ terms. The energy values
are taken from the NIST database \citep{NIST_ASD}. The solid line arrows
indicate the \gls*{E1} transitions, while the dashed line arrows indicate
the \gls*{MIT}s. 
}
\label{fig1}
\end{figure}

The upper ($^7D_{1,2,3,4,5}$) and lower ($^7P^\circ_{2,3,4}$) terms of the
\ion{Cr}{1} \gls*{MIT} at 533.03~nm, as well as the ground level of
\ion{Cr}{1}, are shown in the Grotrian diagram in \fref{fig1}. There are nine
\gls*{E1} transitions between the two terms, indicated by the solid line
arrows. The six transitions indicated by the dashed line arrows are induced due to the
atomic state-mixing in the presence of external magnetic fields, i.e. they
are the \gls*{MIT}s. The corresponding wavelengths are 527.38, 527.46, 
529.66, 529.88, 533.03, and 533.06~nm, respectively.
The energy separations of the fine structure levels of
the upper term $^7D$ are on the order of $10^{-4}$ eV. The small energy
separation allows for a significant degree of state-mixing.

\begin{figure}[htp]
\center
\includegraphics[width=0.50\textwidth]{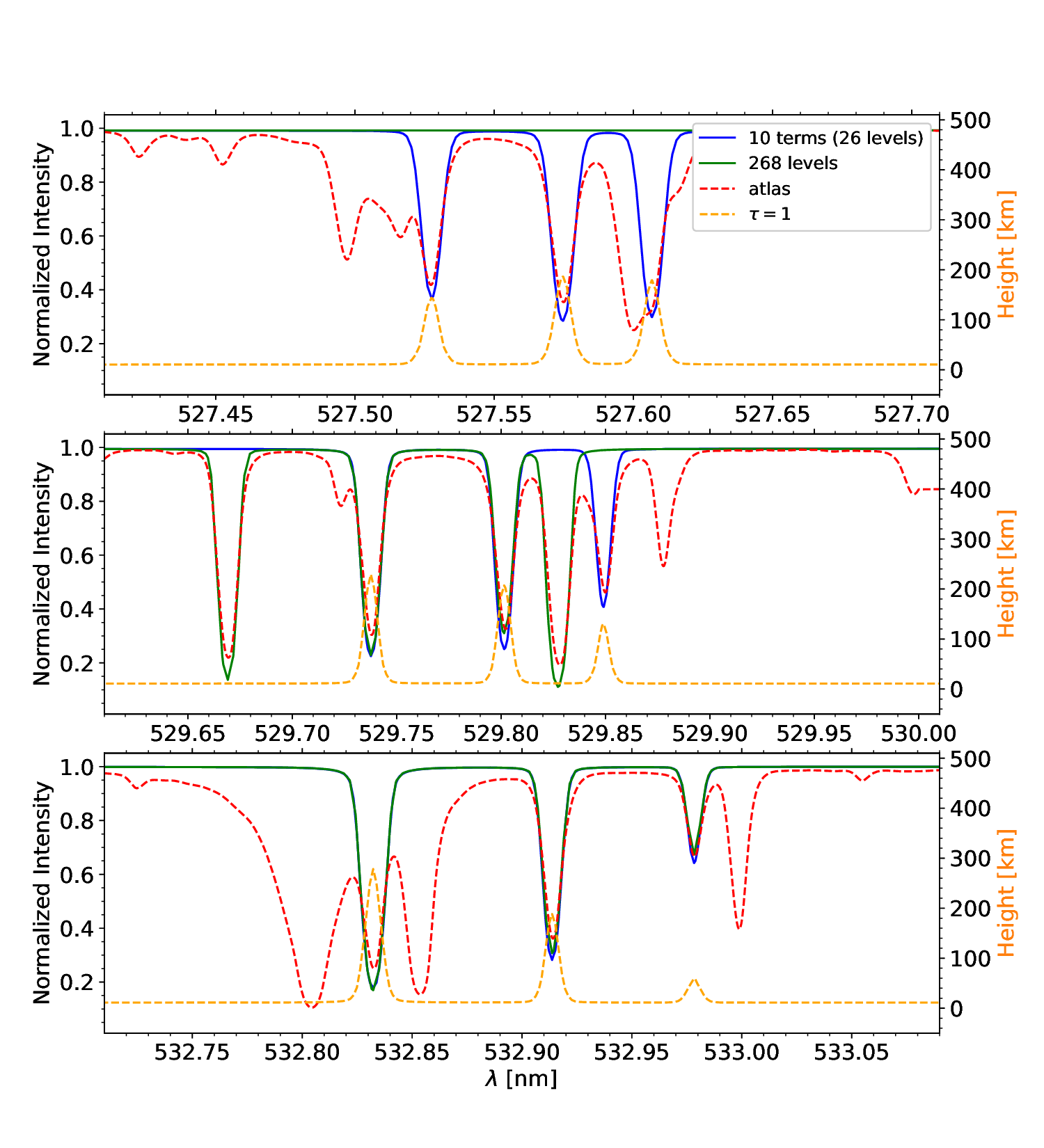}
\caption{Intensity profiles at disk center resulting from the solution of the \gls*{RT}
problem in the FAL-C model with a Cr multi-level model with 268 levels
(green curve) and with a multi-term model with 26 atomic levels (blue curve). 
The red dashed curve shows the solar spectral atlas of the quiet Sun
\citep{Delbouille1973}. The intensity profiles have been normalized to their
value at the continuum. The orange dashed curves (see right axis) show the
height where the optical depth is equal to one. The three panels show
different spectral ranges, covering all spectral lines in the multiplet
of interest.
}
\label{fig2}
\end{figure}

\begin{figure}[htp]
\center
\includegraphics[width=0.5\textwidth]{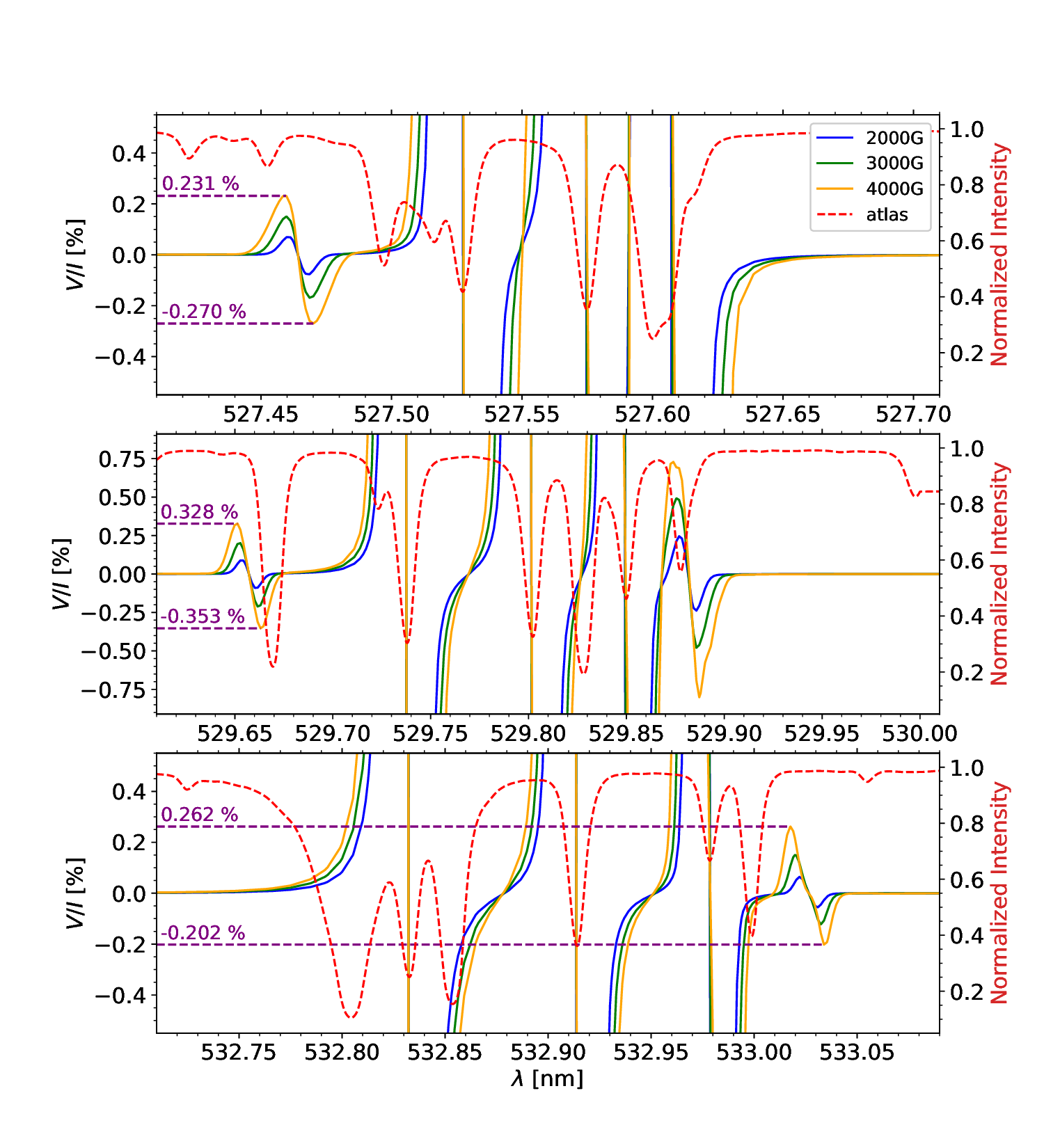}
\caption{Fractional circular polarization $V/I$ profiles at disk center resulting from the
solution of the \gls*{RT} problem in the FAL-C model with a Cr multi-term
model with 26 atomic levels, imposing an homogeneous magnetic field
of 2 (blue curve), 3 (green curve), and 4~kG (orange curve).
The red dashed curve (see right axis) shows the intensity in the solar
spectral atlas of the quiet Sun \citep{Delbouille1973}. The three panels show
different spectral ranges, covering all spectral lines in the multiplet
of interest.
}
\label{fig3}
\end{figure}

\begin{figure}[htp]
\center
\includegraphics[width=0.5\textwidth]{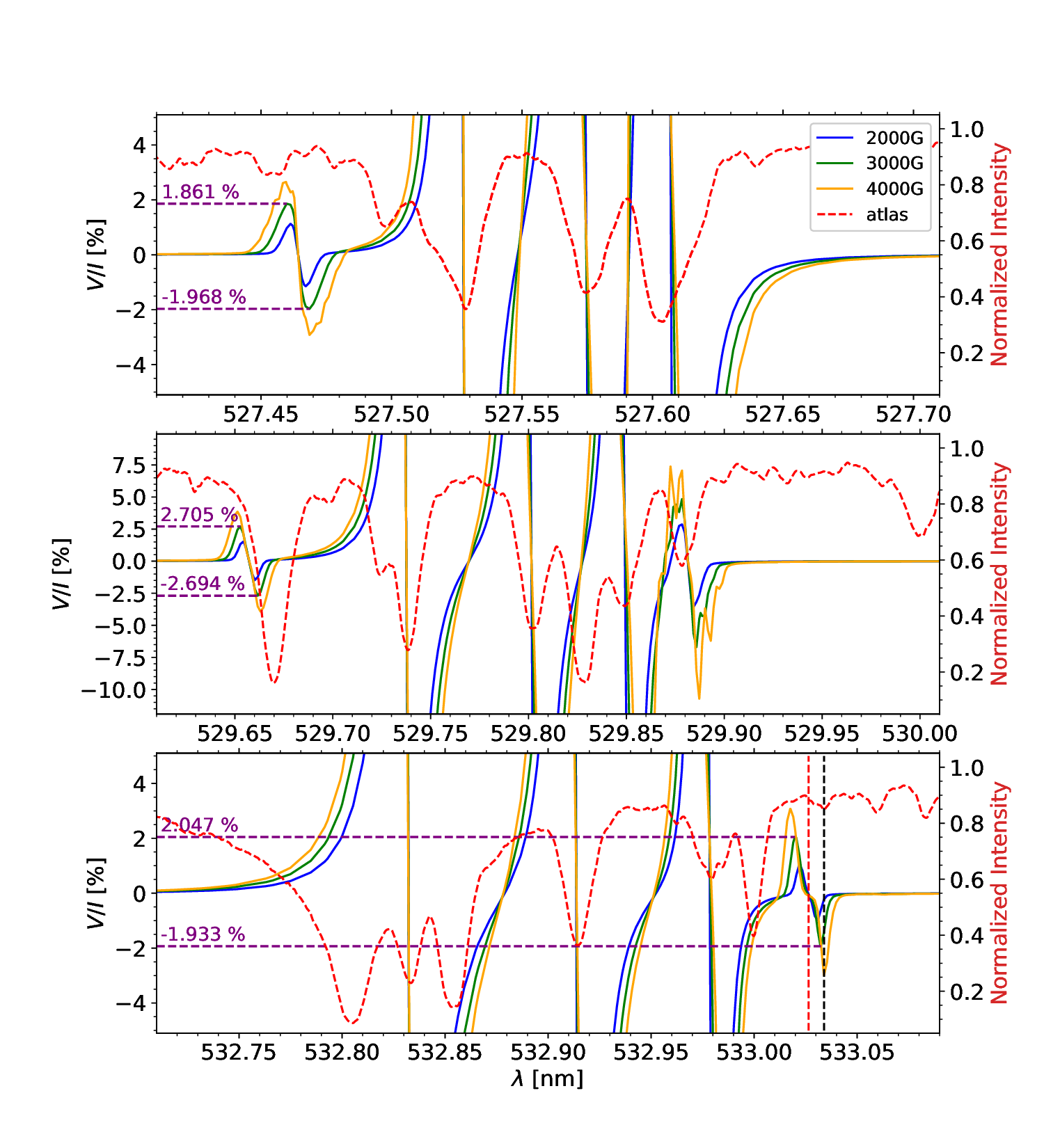}
\caption{Same than \fref{fig3}, but for the M-E model and the sunspot atlas \citep{Wallace2005}.
The vertical black dashed line indicates the center of the absorption feature close to 533.03~nm,
while the vertical red dashed line indicates the zero crossing of the circular polarization
profile of the nearby MIT.}
\label{fig4}
\end{figure}

Within the \gls*{LS} coupling scheme, the transition rate between two terms 
and the rate between two $J$ levels are related by the following
expressions \citep[e.g.,][]{LL04},
\begin{subequations}
\begin{align}
\begin{split}
A(L_uSJ_u\rightarrow L_lSJ_l) & = (2L_u+1)(2J_\ell+1) \\
                                 & \times \begin{Bmatrix} L_u & L_l & 1\\ J_l & J_u & S \end{Bmatrix}^2
                                A(L_u S \rightarrow L_\ell S), \end{split}\label{Eq1a} \\
A(L_u S \rightarrow L_\ell S) & = \sum_{J_\ell}A(L_u S J_u\rightarrow L_\ell SJ_\ell), \label{Eq1b}
\end{align}
\label{Eq1}
\end{subequations}
where $L_u$ and $L_l$ are the orbital angular momenta of the upper and lower
terms, respectively, $J_u$ and $J_l$ are the total angular momenta of the upper
and lower levels, respectively, and $S$ is the spin angular momentum. 
$A(L_uSJ_u\rightarrow L_lSJ_l)$, and $A(L_u S \rightarrow L_\ell S)$ are the 
Einstein coefficients for emission, proportional to the transition rate, from the level $L_uSJ_u$  to 
$L_lSJ_l$, and from the term $L_u S$ to $L_\ell S$, respectively.
In the multi-term approach, the quantum state coupling between the fine structure levels within a term 
(all the $J$ levels for a given spin $S$ and orbital angular momentum $L$) is taken into account, 
and the total angular momentum $J$ is thus not a good quantum number. This coupling is instead 
neglected in the multi-level approach \citep[see \S7 in][for further details on
multi-level and multi-term model atoms]{LL04}.
While in theory the transition probability obtained from Eq.~\eqref{Eq1b} is
expected to be the same for every $J_u$, they often differ in practice. 
It is thus necessary to estimate an average  $A(L_u S \rightarrow L_\ell S)$ accounting 
for every $J_u$ level. In previous works we have taken the average weighted by the degeneracy 
of the upper level \citep{Tanausu2020ApJ}. However, due to the lack of data for some of the fine 
structure transitions in this multiplet, we have taken the average value among the available ones.
\begin{equation}
\begin{split}
\bar{A}(L_u S \rightarrow L_\ell S) = & \frac{1}{5}\sum_{J_uJ_l} 
                                (2L_u+1)^{-1}(2J_\ell+1)^{-1} \\
                                \times &
                                \begin{Bmatrix} L_u & L_l & 1\\ J_l & J_u & S \end{Bmatrix}^{-2}
                                A(L_uSJ_u\rightarrow L_lSJ_l),
\end{split}
\label{Eq1c}
\end{equation}
where the sum runs over the levels for which we have a value of the
transition probability. The $L_uSJ_u \rightarrow L_\ell SJ_\ell$
transition probability which correspond, via Eq.~\eqref{Eq1a},
to the derived $L_uS \rightarrow L_\ell S$ transition probability within the \gls*{LS}
coupling scheme differ from the NIST ones by about a $\sim5$\%, except for the
$J_u=3\rightarrow J_l=3$ transition, which differs by about $\sim20$\%.
In Table \ref{tab1}, we list the transition rates derived via Eq.~\eqref{Eq1a} 
and those from the NIST database.
The energy level of the magnetic sub-levels and the \gls*{MIT} rates as a function 
of magnetic field strength are show in Appendix~\ref{app}.
The \gls*{M2} transition rates between the two terms of interest are on the 
order between $10^{-4}$ and $10^{-7}~s^{-1}$, estimated with the multi-configuration 
Dirac-Hartee-Fock method \citep{Li2020A&A}. These rates are too small to significantly contribute to the 
observable signal, and are therefore neglected.

\begin{table}
  \centering
  \caption{ Wavelengths and transition rates of the nine \gls*{E1} transitions, and 
  the formation heights of the nine lines. $A_{\rm NIST}$ denotes the rate from NIST database, 
  while $A$ is rate obtained from Eq.~\eqref{Eq1a}.}
  \begin{tabular}{c c c c c c} 
  \hline\hline
Transition   &   $\lambda$ (air)   &   $A_{\rm NIST}$  &  $A$           &  $\tau_\lambda=1$ height \\
                   &   (nm)                    &     ($s^{-1}$)          &  ($s^{-1}$)  &  (km) \\

  \hline
  $^7D_3\rightarrow {^7P_2^\circ}$   & 527.53 & -          & 1.67e7            & 140 \\
  $^7D_2\rightarrow {^7P_2^\circ}$   & 527.57 & -          & 3.90e7            & 190 \\
  $^7D_1\rightarrow {^7P_2^\circ}$   & 527.61 & -           & 5.85e7            & 180 \\
  $^7D_4\rightarrow {^7P_3^\circ}$   & 529.74 & 3.88e7 & 3.65e7            & 230 \\
  $^7D_3\rightarrow {^7P_3^\circ}$   & 529.80 & 3.0e7   & 3.65e7            & 210 \\
  $^7D_2\rightarrow {^7P_3^\circ}$   & 529.85 & -           & 1.95e7            & 130 \\
  $^7D_5\rightarrow {^7P_4^\circ}$   & 532.83 & 6.2e7   &  5.85e7           & 280 \\
  $^7D_4\rightarrow {^7P_4^\circ}$   & 532.91 & 2.25e7  & 2.20e7           & 190 \\
  $^7D_3\rightarrow {^7P_4^\circ}$   & 532.98 & 5.38e6  & 5.22e6           & 60 \\
  \hline
  \end{tabular}  \label{tab1}
\end{table}

The bound-bound collisional rates with electrons are estimated 
following \citet{vanRegemorter1962ApJ} for the \gls*{E1} transitions, 
and following \citet{Bely1970ARA&A} for the forbidden transitions.
The bound-free collisional rates with electrons were estimated 
following \citet{Allen1973}. The photoionization cross sections are 
taken as hydrogenic \citep[see, e.g.,][]{Mihalas1978}.

Since the ionization potential of \ion{Cr}{1} is relatively low, we
have some \ion{Cr}{2} levels and the ground level of \ion{Cr}{3} in
our atomic model, as they impact the \ion{Cr}{1} ionization fraction.
We use the HanleRT code \citep{Tanausu2016ApJ,Tanausu2020ApJ} to solve the
\gls*{RT} problem out of local thermodynamical equilibrium (non-LTE).
\fref{fig2} shows the intensity profiles synthesized in the FAL-C model.
The wavelengths are converted from their values in vacuum to those
in air by assuming that the refractive index of air is 1.00027825
at 530~nm \citep{Ciddor1996ApOpt}. The green curves in the figure show
the profiles computed with a multi-level atomic model consisting of 143
\ion{Cr}{1} levels, 122 \ion{Cr}{2} levels, and the ground level of
\ion{Cr}{3}, while the blue curves show the profiles computed with
a multi-term atomic model consisting of 6 (16) \ion{Cr}{1}
terms (levels), 3 (9) \ion{Cr}{2} terms (levels), and the
ground term of \ion{Cr}{3}. The red dashed curves in the figure show
the solar spectral atlas of the quiet Sun
\citep{Delbouille1973}.\footnote{$\rm https://bass2000.obspm.fr/solar\_spect.php$}

Clearly, the 268-level atomic model (green curves) is more accurate,
as the corresponding synthesized profiles can fit the main features
of the Cr lines in the atlas. The Cr absorption lines at
527.53, 527.57, 527.61, and 529.85~nm are missing due to the lack of
transition rates in the NIST database. The blue curves show the profiles
synthesized with a multi-term 10-term (26 levels) atomic model. The
missing transition rates are ``automatically filled'' through the
expressions for the transfer and relaxation rates, and the \gls*{RT}
coefficients (see Eq.~\eqref{Eq1a} and \S7.6 of \citealt{LL04}). The difference between
the synthesis and the spectral atlas is not significant, which indicates
that this reduced model atom is good enough for the purpose of this
paper, significantly reducing the computational requirements. Hereafter,
all the calculations are performed with the 10-term atomic model.

Finally, the orange curves in \fref{fig2} indicate the height where
the optical depth $\tau_\lambda$ is equal to one, which roughly
indicates the formation heights. 
The corresponding heights at the line center are listed in Table \ref{tab1}.
All the nine \gls*{E1} lines form 
in the photosphere, with the line at 532.98~nm forming lower than the rest.

In the presence of external magnetic fields, six transitions are induced 
due to state-mixing. \fref{fig3} shows the Stokes $V/I$ profiles 
synthesized in the FAL-C model with longitudinal magnetic fields of 
2000, 3000, and 4000~G, respectively, for a line of sight with $\mu = 1.0$, 
where $\mu$ is the cosine of the heliocentric angle. The $V/I$ amplitudes of the 
four \gls*{MIT}s are on the order of $10^{-3}$. The other two \gls*{MIT}s 
are not shown due to their weak signal. The amplitude of the line at 
529.88~nm is larger than those of the other 3 lines, since the energy 
separation between $^7D_1$ and $^7D_2$ is smaller than others. 
Besides, it can be seen that the amplitudes of the two lobes of the 
\gls*{MIT} are not exactly the same. This is because the \gls*{MIT} 
rates between M states are not symmetric due to 
the different degrees of state-mixing for each $M$ state (see Appendix~\ref{app}).
The asymmetric nature of these transition rates can also be seen in
Table 3 in \citet{Li2021ApJ}, and are only significant for strong magnetic fields.  

Figure~\ref{fig4} shows the synthesized profiles in the M-E model. 
In the sunspot model, the temperature is lower than in the FAL-C model.
More Cr atoms are found in its neutral stage and the lines are narrower. 
Consequently, larger $V/I$ amplitudes of around $2\%$ in the \gls*{MIT} lines 
are present. The profiles at 527.46~nm and 529.88~nm are not smooth for a field 
strength of 4000~G, because the Zeeman splitting of such a strong magnetic field is 
comparable with the line width. The red dashed curves show the spectral atlas of a sunspot 
\citep{Wallace2005}.\footnote{$\rm https://nispdata.nso.edu/ftp/pub/atlas/spot4atl/$}
In contrast to the atlas of the quiet sun, there is an additional weak absorption 
feature near 533.03~nm. 
The center of the absorption is denoted by the black dashed curve, and it does not
coincide with the zero crossing of the circular polarization profile of the \gls*{MIT}
(indicated with a red dashed line in the figure). This absorption feature
could correspond with an unknown molecular line.
Comparing with the atlas, we find that the two of the \gls*{MIT} lines at 
529.65 and 529.87~nm are blended with strong lines. 
Contrarily, the lines at 527.46 and 533.03~nm are not too much contaminated.
Especially, the line at 533.03~nm appears to blend solely with an unknown weak absorption feature, 
and it should be detectable with current solar telescopes and instrumentation.

\section{Observation and inversion} \label{sec:intro}

\begin{figure*}[htp]
\center
\includegraphics[width=0.8\textwidth]{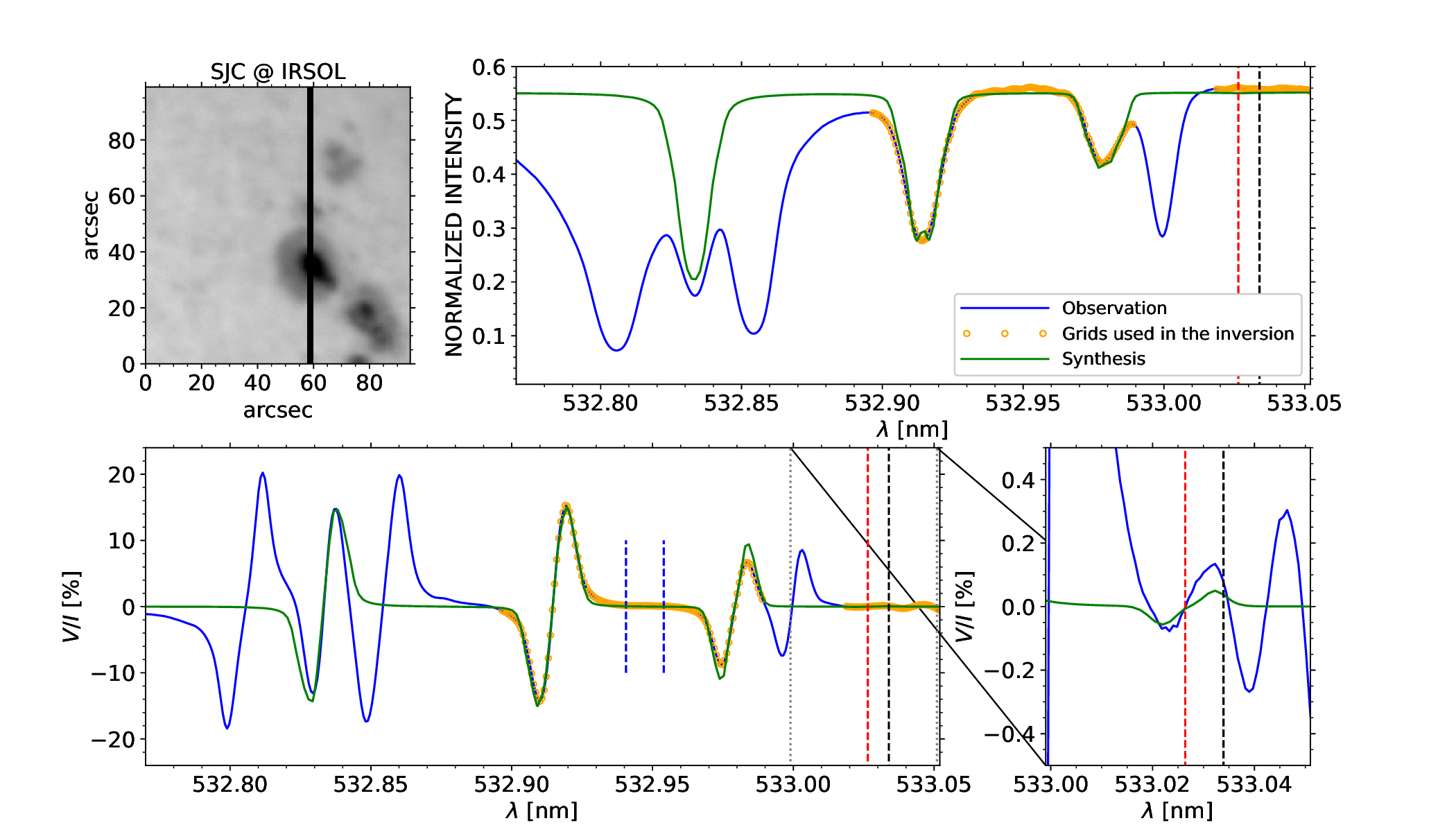}
\caption{Slit-jaw image (top left), intensity profile (top right), and
fractional circular polarization $V/I$ profile (bottom row) for the NOAA
AR 13102 observed on 20 September 2022 with the ZIMPOL-3 at the \gls*{IRSOL} observatory.
The black line in the top left panel indicates the location of the
spectrograph's slit. The blue solid curves show the profiles in the
observation, the green solid curves show the profiles in the synthesis
in the atmospheric model resulting from the inversion, and the orange
open circles indicate the wavelengths included in the inversion. The
red dashed line indicates the line center of the \gls*{MIT} and the black
dashed line indicates the center of an absorption feature in the sunspot
atlas. Finally, the dotted lines indicate the wavelength range plotted
in the bottom right panel. The noise level is 0.023\% estimated 
by the \gls*{RMS} of the fractional circular polarization $V/I$ in the
continuum between the two vertical blue dashed lines.}
\label{fig5}
\end{figure*}

\begin{figure*}[htp]
\center
\includegraphics[width=0.8\textwidth]{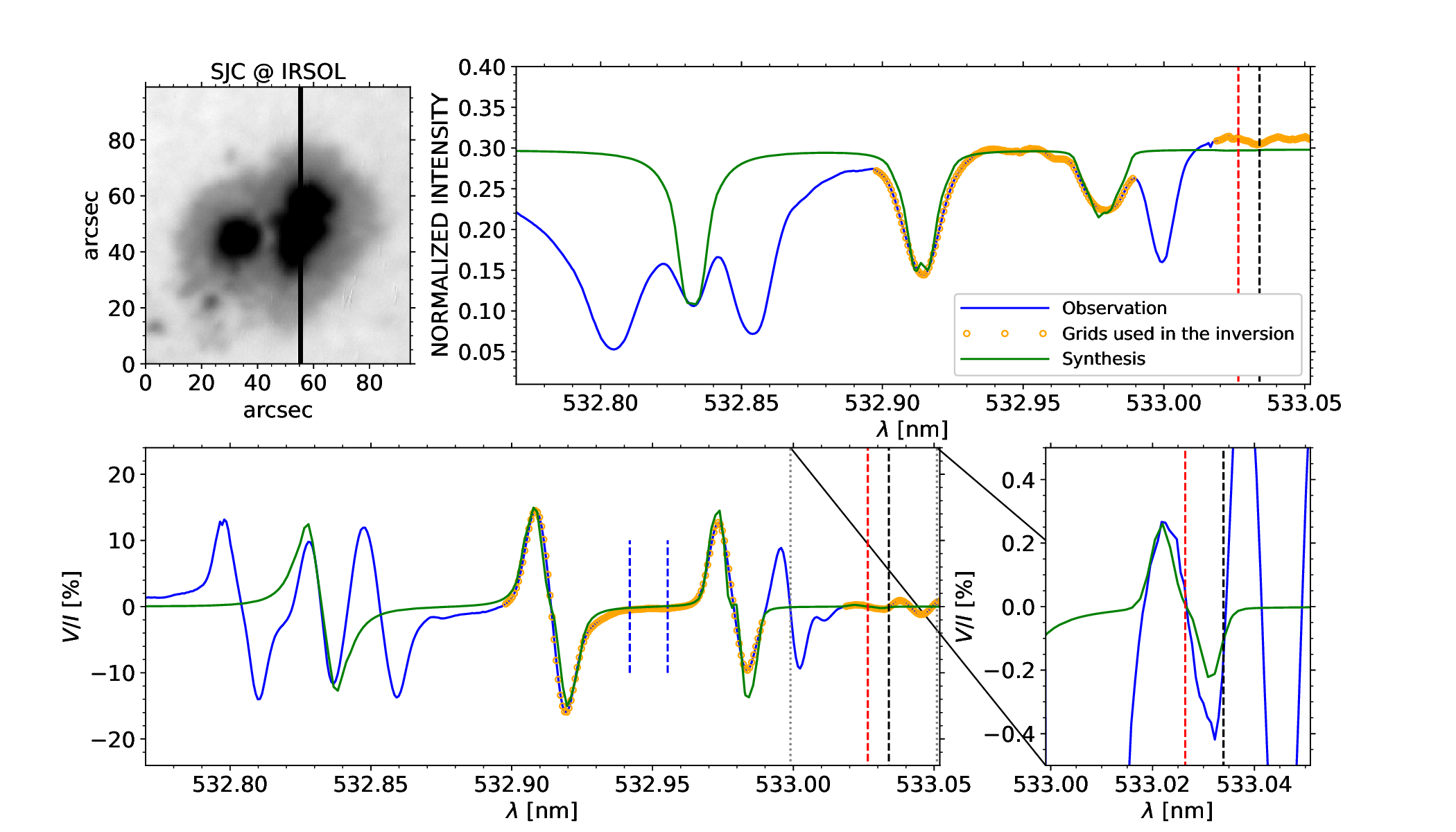}
\caption{Same as \fref{fig5}, but for NOAA AR 13153 observed on 6 December 2022.
The noise level is 0.068\% estimated by the \gls*{RMS} of the $V/I$ in the
continuum between the two vertical blue dashed lines.}
\label{fig6}
\end{figure*}

Spectro-polarimetric observations of the intensity and circular polarization
of the \ion{Cr}{1} lines around 533~nm in the NOAA AR 13102 and
13153 were acquired on 20 September and 6 December 2022, respectively.
Both were located far from the solar limb. We employed the high-sensitivity 
ZIMPOL-3 spectropolarimeter (using the photo elastic modulator, as described 
in \citealt{Ramelli2010SPIE}) at the 45 cm aperture IRSOL Gregory Coud\'e 
Telescope in Locarno, Switzerland, alongside a Czerny-Turner spectrograph, 
incorporating an interference pre-filter. The spectral sampling is 7.5 m\AA. 
The slit-jaw images are shown in the top left panels of Figs.~\ref{fig5} and \ref{fig6}, 
with the position of the slit indicated by the black line. 
Additionally, the slow modulation technique was used to enhance the zero-level 
accuracy of the circular polarization \citep{Zeuner2022SPIE}. The data are reduced 
with the standard reduction (dark image and flat field correction as well as 
polarimetric calibration) software but are not corrected for spatial or spectral stray light. 
The spectral stray light is on the order of 2\%, mostly affecting the depth of the spectral 
lines. The core of the sunspot might be compromised by stray light contributions from 
the telescope, instruments, and the sky. As a result, the Stokes $V/I$ profile amplitudes 
might be decreased. However, as we will show later, we are not interested in a detailed 
inversion of this region, as our argument is based on a consistent modeling of multiple 
Stokes $V/I$ profiles.
The intensity and fractional circular polarization
profiles resulting from the spatial average over several arcsec along the spectrograph
slit is shown in the top right and bottom row of Figs.~\ref{fig5} and \ref{fig6}, respectively.
In the bottom right panel of these figures
we show the fractional circular polarization $V/I$ around the \gls*{MIT}. The
red dashed line indicates the line center of the \gls*{MIT}, while the
black dashed line indicates the line center of the absorption feature observed
in the sunspot atlas. There is no evidence of any feature in the intensity
profile that can be identified as the \gls*{MIT}. However, the fractional circular
polarization profile shows an approximately anti-symmetric shape centered in
the wavelength where the \gls*{MIT} is expected. The amplitude of this
fractional circular polarization is about 0.1\% in NOAA AR 13102 and 0.2\% in NOAA
AR 13153, clearly above the statistical noise level of 0.02\% and 0.07\% estimated by the 
\gls*{RMS} of the fractional circular polarization $V/I$ profile in the continuum
between the two blue vertical lines shown in the bottom left panel of
Figs.~\ref{fig5} and \ref{fig6}.

We applied our non-LTE inversion code, HanleRT-TIC, to the observations of the
\ion{Cr}{1} permitted lines at 532.91 and 532.98~nm, and the \gls*{MIT} at
533.03~nm, assuming $\mu = 1.0$. 
The orange circles in Figs.~\ref{fig5} and \ref{fig6} indicate the wavelengths 
included in the inversion. We considered a model atmosphere stratified in
60 nodes distributed between
\TAUE{-6.5} and 1.5, and gave a larger weight (50 times larger than the E1 transitions) 
to the circular polarization of the \gls*{MIT} under investigation due to its weak signal. 
Two cycles are employed to invert the profiles. The gas pressure at the top boundary is fixed 
at 1~dyn/cm$^{-2}$. In the first cycle, we adopted 4 nodes in the 
field strength, 4 nodes in the inclination of the magnetic field, 5 nodes in the temperature, 
and 1 node in the vertical velocity. Note that we did not observe the linear polarization. 
The transverse magnetic field is included as a source of line broadening. Besides it 
also contributes to the level mixing and furthermore impacts the \gls*{MIT}. 
In the second cycle, we adopted 4 nodes in the field strength, 4 nodes in the inclination 
of the magnetic field, 7 nodes in the temperature, 4 nodes in the vertical velocity, and 1 
node in the micro-turbulent velocity to improve the fitting.
The inverted longitudinal magnetic
field in the lower photosphere of the inverted models is about -1000~G for
NOAA AR 13102 and about 2000~G for NOAA AR 13153.

The green curve in Figs.~\ref{fig5} and \ref{fig6} show the intensity and fractional
circular polarization profiles synthesized in the inferred model atmosphere.
The synthetic profiles reproduce the main features of the observation, even for the
line at 532.83~nm despite it being neglected in the inversion. The fractional
circular polarization amplitudes of the \gls*{MIT} are similar to those in the
observed profiles. The main reason for the differences are likely due to the
fact that the observed profiles are averages over several arcsec along the slit, 
mixing different parts of the sunspot, which cannot be represented in a 
single and relatively simple sunspot model.
Nearby lines might also contaminate the profiles, especially the probably unknown 
molecular line indicated by the black dashed line in \fref{fig4}-\ref{fig6}.

\section{Conclusion} \label{sec:intro}

In this work we have identified a \ion{Cr}{1} \gls*{MIT} at 533.03~nm. The upper term, 
$3d^5(^6S)4d\ ^7D$, consists of 5 fine structure levels with an energy separation
on the order of $\rm 10^{-4}$~eV, a relatively small energy which facilitates
the $J$-state mixing in the presence of external magnetic fields with a strength
that can be typically found in the solar photosphere of solar active regions. 
The solution of the non-LTE \gls*{RT} problem in semi-empirical models shows
fractional circular polarization $V/I$ amplitudes of about $0.1$\% in a quiet
Sun model and about $1$\% in a sunspot model, when imposing a $3000$~G longitudinal
magnetic field in both cases. The amplitudes of the circular polarization lobes 
of the \gls*{MIT} are not exactly anti-symmetrical, due to the different 
degrees of state-mixing for each $M$ state.

Spectro-polarimetric observations of two active regions were obtained with the
ZIMPOL-3 at the \gls*{IRSOL}. While there is no feature in the intensity profile
that can be identified with the \ion{Cr}{1} \gls*{MIT}, the fractional circular
polarization profiles show a clear approximately anti-symmetric shape whose center
coincides with the wavelength predicted for the \gls*{MIT}. The amplitudes of the
circular polarization profiles are about $0.1$\% and $0.2$\% in the two sunspots,
well above the polarization noise level estimated from the 
\gls*{RMS} of the fractional circular polarization $V/I$ in the continuum.

Applying HanleRT-TIC, we performed non-LTE inversions on the profiles of three \ion{Cr}{1} lines in the
observation, including the \gls*{MIT}. We show the intensity and circular polarization
profiles resulting from the solution of the non-LTE \gls*{RT} problem in the
inferred model atmospheres. These synthetic profiles show fractional
circular polarization amplitudes similar to those in the observations, with the
discrepancies likely explained by the contamination of nearby lines (partial blends)
and the spatial averaging of the data over several arcsec. All-in-all, the fractional
circular polarization profiles observed at 533.03~nm, with no counterpart in the
intensity profile, are very likely due to the predicted \ion{Cr}{1} \gls*{MIT}.

While \gls*{MIT} lines such as the one studied in this paper are only detectable in the
presence of magnetic fields above a certain strength, the potential for photospheric
magnetic field diagnostic of these lines is extremely limited. While the mere
presence of the \gls*{MIT} line is evidence of a magnetic field, the magnetic field
strength that is usually required is so large that the magnetic field is already
evident from several other observables (e.g., the sunspot itself). Moreover, due
to the small strength expected for the \gls*{MIT} for the typical magnetic fields
expected in the Sun, their signals are very susceptible even to partial blends
with nearby spectral lines. Therefore, the usefulness of the \gls*{MIT} for the
diagnosis of solar magnetic fields is likely limited to highly ionized species in
the corona.

\acknowledgements
We thank Michele Bianda for helpful technical assistance,  Andr\'{e}s Asensio 
Ramos for helpful comments and suggestions, and Wenxian Li for carefully reading the 
manuscript and providing us the M2 transition rates.
We acknowledge the funding received from the European Research Council (ERC) 
under the European Union's Horizon 2020 research and innovation programme
(ERC Advanced Grant agreement No 742265). 
T.P.A.'s participation in the publication is part of the Project RYC2021-034006-I, 
funded by MICIN/AEI/10.13039/501100011033, and the European Union 
“NextGenerationEU”/RTRP. 
T.P.A. and J.T.B. acknowledge support from the Agencia Estatal de Investigación del 
Ministerio de Ciencia, Innovación y Universidades (MCIU/AEI) under grant 
``Polarimetric Inference of Magnetic Fields'' and the European Regional Development 
Fund (ERDF) with reference PID2022-136563NB-I00/10.13039/501100011033.
F.Z. acknowledges funding from the European’s Horizon 2020 research and innovation 
programme under grant agreement no 824135, and the Swiss National Science 
Foundation under grant number 200020\_213147. IRSOL is supported by the Swiss 
Confederation (SEFRI),  Canton Ticino, the city of Locarno and the local municipalities.

\appendix
\counterwithin{figure}{section}

\section{The level energy of the magnetic sub-levels and the \gls*{MIT} rates}\label{app}

\begin{figure*}[htp]
\center
\includegraphics[width=1.0\textwidth]{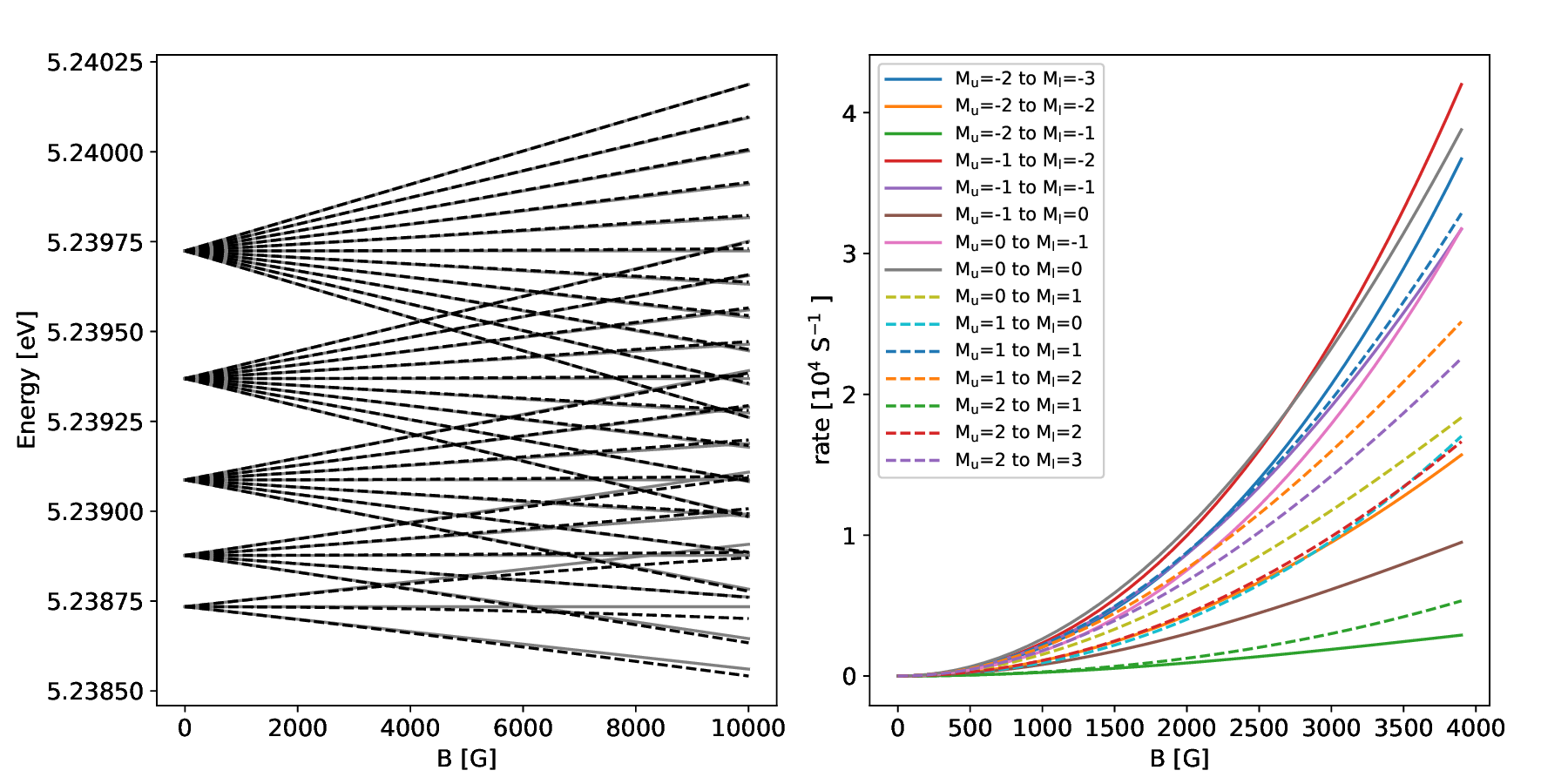}
\caption{Left panel: Black dashed curves show the level energy of the $3d^5(^6S)4d\ ^7D$ 
magnetic sublevels as a function of the magnetic field strength.
The gray solid curves show the same energy when assuming the linear Zeeman 
approximation (neglecting energy coupling).
Right panel: The \gls*{MIT} transition rates between $3d^5(^6S)4d\ ^7D_2$ and 
$3d^5(^6S)4p\ ^7P_4^\circ$ states as a function of the magnetic field strength.
}
\label{figa1}
\end{figure*}

The level energy of the upper term $3d^5(^6S)4d\ ^7D$ as a function of field strength shown 
in the left panel of \fref{figa1} are computed by diagonalizing the atomic Hamiltonian
according to Eqs.~(3.61a) and (3.61b) in \citet{LL04}. 
The \gls*{MIT} rates shown in the right panel of the figure are estimated from Eq.~(7.34b) in
\citet{LL04}, with $jM = j'M'$ and $j_uM_u=j'_u M'_u$.
The rates corresponding to a magnetic field strength of about 3000~G are sufficiently large
as to be able to neglect the contribution of the \gls*{M2} transition rate, of
about $\rm 2.29\times10^{-6}\ S^{-1}$, estimated with the multi-configuration Dirac-Hartee-Fock 
method \citep{Li2020A&A}. 
The lower term $3d^5(^6S)4p ^7P^\circ$ consists of three levels with $J=2$, $3$, and $4$, respectively. 
The fine structure separation is on the order of $10^{-2}$ eV, which is much larger than 
the Zeeman splitting for the typical field strengths in the solar photosphere. 
Therefore, the level coupling is negligible, and the magnetic splitting is still linear.

\bibliography{MIT.bib}{}
\bibliographystyle{aasjournal}



\end{document}